\documentclass[11pt,a4paper]{scrartcl}

\usepackage{ILD}

\usepackage{multirow}
\usepackage[symbol]{footmisc}
\usepackage{feynmf}
\usepackage{textpos}
\usepackage{lineno}

\title{Status of ILD new 250 GeV common MC sample production}

\ildproc{soft}{2021}{002}

\date{\today}

\addauthor{Hiroaki Ono}{\institute{1}}
\addauthor{Akiya Miyamoto}{\institute{2}}

\addinstitute{1}{Nippon Dental University School of Life Dentistry at Niigata}
\addinstitute{2}{High Energy Accelerator Research Organization (KEK)}

\abstract{
We are producing high statistics 250 GeV common MC samples 
for the ILD physics study using the latest generator, simulation, and reconstruction packages.
In the production, 
we utilize ILCDirac distributed computing environment.
The estimated resource requirements, the current status and prospects of 
the production are presented.
}

\titlecomment{Talk presented at the International Workshop on Future Linear Colliders (LCWS2021), \\15-18 March 2021. C21-03-15.1.}

\graphicspath{ {./logos/}{./figures/} }

\begin{document}
\titlepage


 \section{Introduction}
 Two large common Monte Carlo (MC) simulation samples were produced for the study of the ILD detector concept, which was established in 2009~\cite{ILD}. 
One was used for the ILC physics case study as ILC DBD~\cite{DBD} with center of mass (CM) energies of 250 GeV, 350 GeV, 500 GeV, and 1 TeV. The other was produced for the ILD detector optimization study, ILD IDR~\cite{IDR},  using the DBD generator samples at the CM energy of 500 GeV in 2018.\\
~~Since generator, simulation and reconstruction tools are well advanced since the DBD era, 
a new full set of high statistics MC samples with the latest software, named MC-2020,  are requested from the physics working group for precise studies at the CM energy of 250 GeV.  
Table~\ref{tab:request_statistics} shows the summary of requested statistics of MC samples from ILD physics working group.
In this request, $1~\mathrm{ab^{-1}}$ or $5~\mathrm{ ab^{-1}}$ are required for most of channels and $100~\mathrm{ k~events}$ for each of individual higgs decay channels. It is more than an order larger than the previously produced samples.
In order to meet the request of this huge production, we employ the DIRAC~\cite{DIRAC} and ILCDirac~\cite{ILCDIRAC} distributed computing system for mass production which was fully utilized in the ILD group since IDR production period. 
DIRAC is originally developed for the LHCb experiment~\cite{LHCb} providing high level interface for users to use heterogeneous GRID resources, composed of job management, file catalog, job monitoring, control via web interface. ILCDirac is an extension of the DIRAC for linear collider experiments which is developed and maintained by the CLICdp group~\cite{CLIC}.

\begin{table}[htbp]
\begin{center}
\caption{Requested integrated luminosity or number of events for each process. Here 
($e_{L}.p_{R}$, $e_{R}.p_{L}$, $e_{L}.p_{L}$, $e_{R}.p_{R}$) denote 100\% polarized (Left/Right, Right/Left, Left/Left, Right/Right)-handed electron/positron beams.}
\label{tab:request_statistics}
\begin{tabular}{|c|c|}
\hline
Process & Statistics \\
\hline
\hline
\multirow{2}{*}{$2f\to \ell\ell,~ qq$} & $e_L.p_R/e_R.p_L$ : $\mathrm{ 5~ab^{-1}}$\\
& $e_L.p_L/e_R.p_R$ : $\mathrm{1~ab^{-1}}$\\
\hline
\multirow{2}{*}{All $4f$} & $e_L.p_R/e_R.p_L$ : $\mathrm{ 5~ab^{-1}}$\\
& $e_L.p_L/e_R.p_R$ : $\mathrm{ 1~ab^{-1}}$\\
\hline
All $6f$ & $\mathrm{ 10~k~events}$\\
\hline
$e\gamma$, $\gamma e$, $\gamma\gamma$ process ($3f$, $5f$, $\gamma\gamma\to 2f$, $\gamma\gamma \to 4f$) & $\mathrm{1~ab^{-1}}$\\
\hline
$2f \to ee\gamma$ & $\mathrm{1~ab^{-1}}$\\
\hline
$h\to$ inclusive & $\mathrm{1~ab^{-1}}$\\
\hline
$h\to$ each decay ($5\times 9$ channels) & $\mathrm{100~k~events}$\\
\hline
$Z \to qq$, $Zh\to \nu\nu qq$ for LCFI & $\mathrm{50~k~events}$\\
\hline
$Z\to qq$ (91 GeV) for LCFI & $\mathrm{50~k~events}$\\
\hline
\end{tabular}
\end{center}
\end{table}

For MC-2020, ILD generator group produces the event samples using the latest Whizard 2.8.5~\cite{Whizard}.
In this production, high statistics generator samples are newly produced even with small cross section channels.
With the latest Whizard, generator outputs are directly stored in the LCIO~\cite{LCIO} format instead of previous stdhep format in the DBD sample. 
The generated samples are simulated with the ILD large detector model (ILD\_l5\_v02) using the dd4hep~\cite{dd4hep}.
ILD\_l5\_v02 is a hybrid calorimeter model composed of the silicon and scintillator tungsten electromagnetic calorimeter and analog scintillator and semi-digital RPC hadron calorimeter. 
The simulated data (SIM) are reconstructed by the Marlin package~\cite{Marlin} with the silicon electromagnetic and analog scintillator hadron calorimeter option (ILD\_l5\_o1\_v02), creating REC and DST files. 
REC files contain full reconstruction information while DST files contain only partial information suitable for physics studies. 
The software version is ILCSoft-v02-02~\cite{ILCSoft}.
In the reconstruction process, low $p_T$ hadrons by $\gamma\gamma$ interactions and $e^{+}e^{-}$ pairs by bremsstrahlung backgrounds are overlaid to the simulated events.
According to an estimation, required storage space for the requested statistics is $4.6~\mathrm{PB}$ by SIM and $5.5~\mathrm{PB}$ by REC, which exceeds the available storage capacity of ILD.
Thus we decided not to keep all the SIM and REC files.
In this production, DST files are only all kept but only 10\% or at least 500 jobs of REC files are partially kept.
Similar rule applied to the SIM files.
 
\section{MC mass production workflow in ILD}
\begin{figure}[htbp]
\begin{center}
 \includegraphics[width=0.5\textwidth]{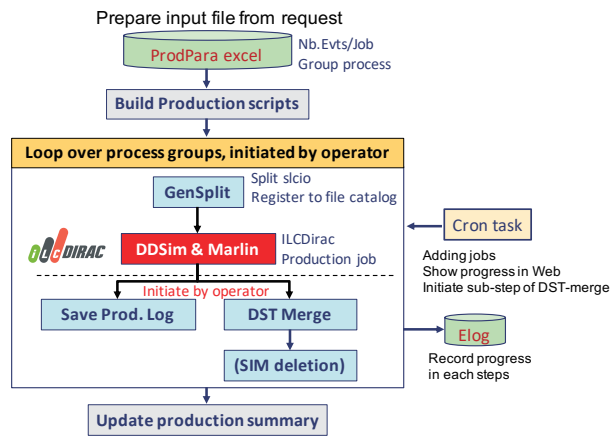}
 \caption{ILD production workflow with ILCDirac managed by ILDProd tool.}
 \label{Fig:workflow}
 \end{center}
 \end{figure}

ILD MC production workflow is conducted by the ILDProd tool~\cite{ILDProd} utilizing ILCDirac as shown in Fig.~{\ref{Fig:workflow}}.
At first, a list of production request is prepared in an excel format.
Generator file splitter and production transformation submission scripts are created from the input excel file, 
and a json format file is created for controlling subsequent production workflow by python, 
then a series of production is launched by a start script by operator.
After launching production, 
at the first step, generator samples are split into small chunk by GenSplit. It is processed at the KEKCC local system.
Split generator files are then uploaded to the KEK-DISK storage element (SE), registered into a DIRAC file catalog, 
and meta information is attached.
ILCDirac production transformations are submitted for simulation and reconstruction soon after launching the GenSplit process.
In the MC-2020 production, following ILCDirac transformations are used in order to cope with the difference of keeping files: 
SIM, REC partially-kept, DST only-kept.  REC and DST files are fully kept in the case of calibration samples.
Job submission and rescheduling in the case of job failure are managed by ILCDirac. Produced files are uploaded to KEK or DESY disk only SEs.
In order to reduce the heavy load by retrieving the background files for overlay,
$\gamma\gamma$ low $p_T$ hadron and $e^{+}e^{-}$ pair background samples are replicated to KEK, DESY, and CERN SEs.\\
~~The completion of transformation is decided by the operator when all input files without error are processed.
After terminating the production transformations, DST merge and log retrieve steps are initiated by operator.
Since DST file size is relatively small, merged-DST files are created by merging 50 to 100 DST files. 
The DST merge is done by utilizing ILCDirac user job and rescheduling in the case of job failure is managed by the ILDProd tool.
All job output logs are also retrieved and archived on the SEs complying the ILD production convention.
Production status is monitored via a web-based monitor and jobs are added periodically with a cron task.
In MC-2020 production, an automation  of rescheduling of DST-merge jobs and retrying of GenSplit registration step
is enhanced by ILDProd, and a need of manual intervention of production process is reduced compared to the previous production.
All the production step statuses are recorded automatically on ELOG~\cite{ELOG} system and
a production summary page~\cite{ProdSummary} is updated automatically when the DST merge and log retrieve 
steps are completed.
This workflow is repeated until all the input groups are processed.

 \section{MC production progress and status}

\begin{figure}[htbp]
 \begin{center}
 \includegraphics[width=0.5\textwidth]{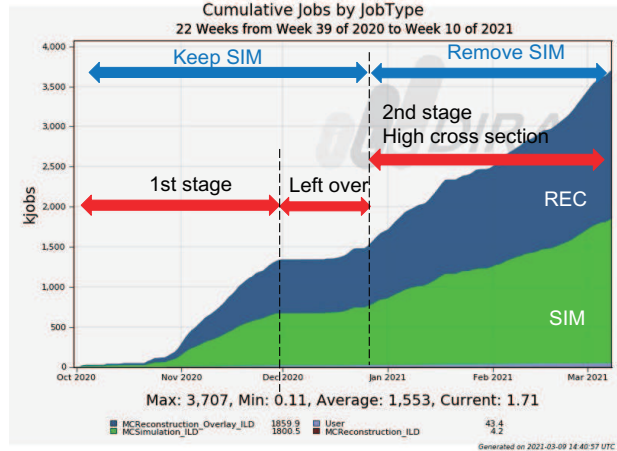}
 \caption{Accumulated number of production jobs for simulated (SIM) and reconstructed (REC) production.}
\label{Fig:ProdProgress}
\end{center}
 \end{figure}

In October 2020, MC-2020 production has been launched
and single particle, higgs, and small or middle cross section channels were processed by the end of November 2020
as ``1st-stage'' period shown in Fig.~\ref{Fig:ProdProgress}.
After this period, some failure jobs, and LCFIPlus~\cite{LCFIPlus} vertex finder training samples were processed as of  ``left-over'' period.
In these period, all the SIM files were kept. In the ``2nd-stage'' period, 
SIM files were not kept any more and they were removed after the completion of merged-DST production.
In this period, more than 3.5 M jobs have been processed and process is still on-going.

\begin{figure}[htbp]
 \begin{minipage}{0.48\textwidth}
 \begin{center}
 \includegraphics[width=0.9\textwidth]{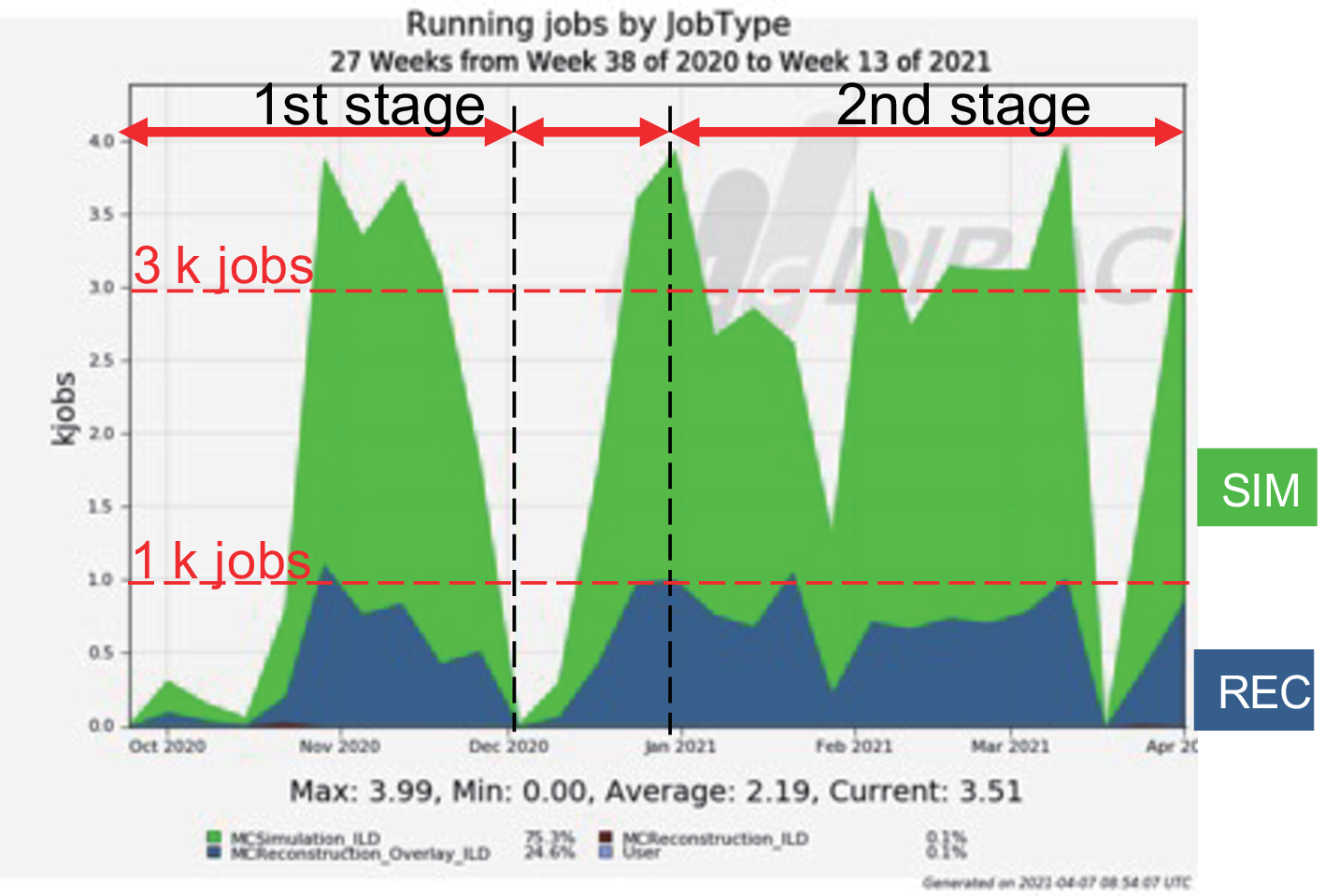}
 \caption{Executed production jobs by job types.}
 \label{Fig:JobType}
  \end{center}
 \end{minipage}
  \begin{minipage}{0.48\textwidth}
  \begin{center}
 \includegraphics[width=0.9\textwidth]{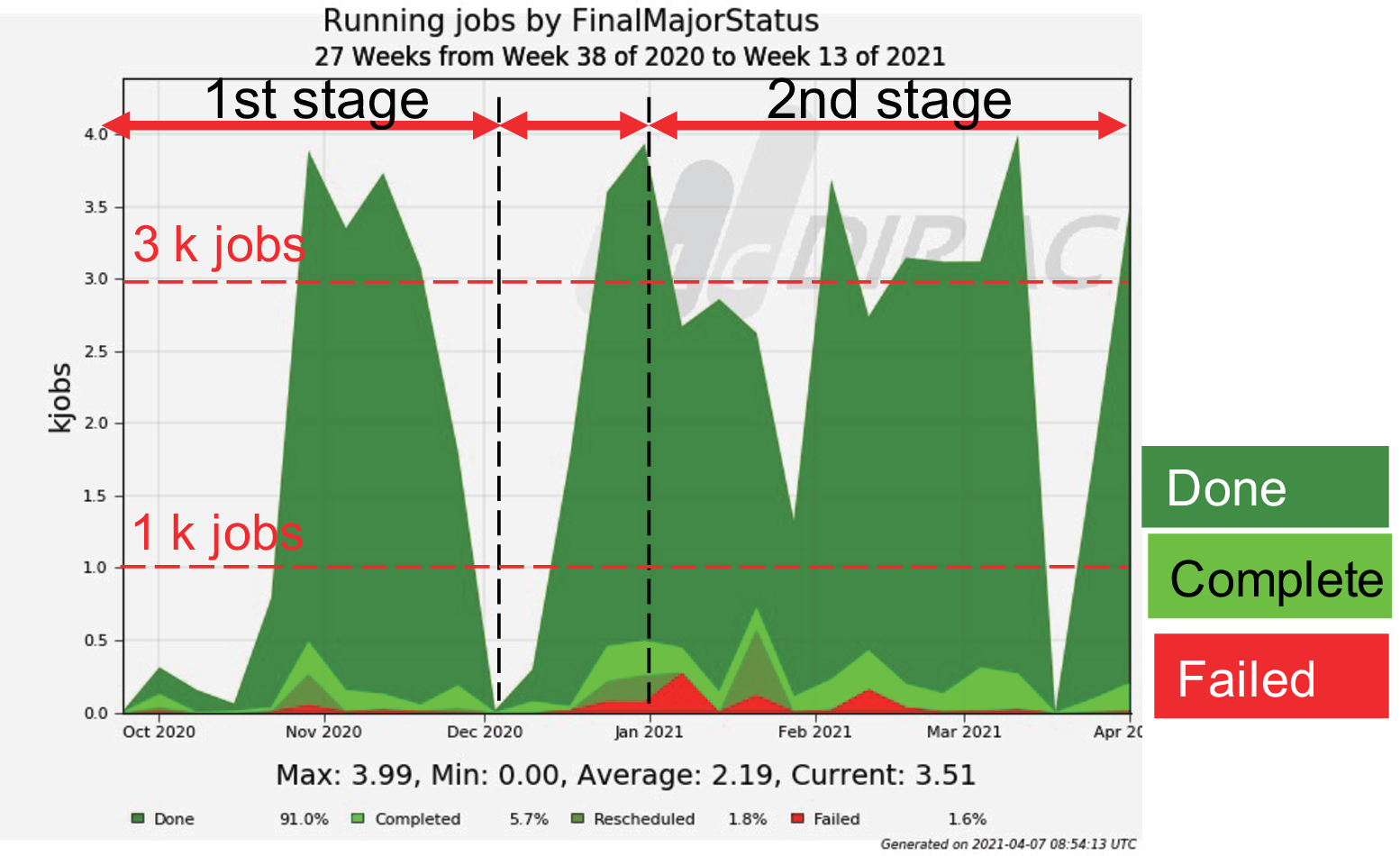}
 \caption{Executed production jobs by final statuses.}
 \label{Fig:JobStatus}
 \end{center}
 \end{minipage}
 \end{figure}

\begin{figure}[htbp]
\begin{center}
 \includegraphics[width=0.5\textwidth]{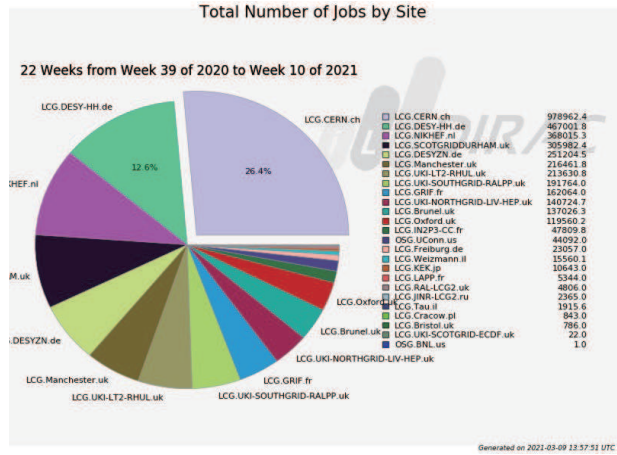}
 \caption{Summary of the job executed sites.}
 \label{Fig:JobSites}
 \end{center}
 \end{figure}

 Figure~\ref{Fig:JobType} and \ref{Fig:JobStatus} show the number of processed jobs clustered by job types and final status. 
 In the period of this figure, 3 to 4 k jobs run concurrently in average, about 1 k of them were REC jobs.
 At the early stage of the production, we faced high job failure rate by ``Stalling'' of REC jobs,  
 which could be caused by the slow or timeout of downloading background overlay files.
 Even though we replicated BG overlay files to several SEs, we found that jobs tried to download BG overlay file only from KEK-DISK SE, resulting high network overload.
 Figure~\ref{Fig:JobSites} shows the job executed sites of this period, showing that jobs were mostly processed at European sites.
Therefore, we were forced to limit the number of concurrent running REC jobs up to 1.5 k in this period.

 The problem was notified to the ILCDirac administrator through JIRA issue tracking system.
 Thanks to the ILCDirac administrators effort, an implementation of SE selection mechanism was improved in the ILCDirac.
 Also we restricted the SE for output files to DESY-SRM in order to avoid the heavy load to KEK-DISK.
 After applying these treatment, high failure rate of the REC overlay process has been reduced and productions are much stabilized.
 
\begin{figure}[htbp]
 \begin{center}
 \includegraphics[width=0.5\textwidth]{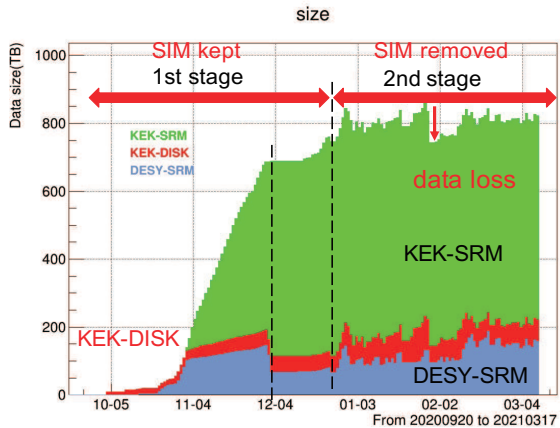}
 \caption{Occupancy of storage space used in the ILD production; KEK-DISK, KEK-SRM, and DESY-SRM.}
 \label{Fig:StorageSpace}
 \end{center}
 \end{figure}

 Figure~\ref{Fig:StorageSpace} shows the occupancy of the storage used in the ILD production.
 A capacity of the disk storage is 300 TB for KEK-DISK SE and 600 TB for DESY-SRM.
 SIM files, which were stored in disk storage in early period were moved to the KEK-SRM (TAPE)  SE 
 and removed from disk storage in a timely manner.
 Unfortunately, due to a miss operation by a production team at the end of January 2021, 
 SIM and REC files were removed, but merged-DST files were not affected.
The lost files included background files for overlay.  The production were restarted after 
 reproducing background files. The 2nd stage $4f$ and $2f$ channels were produced 
 in this period and it continued rather smoothly. 
 
 At the middle of March 2021,  ILCDirac and its base DIRAC version have been updated on the ILCDirac server.
 For this update, production jobs have been drained before the downtime and production was paused.
 After resuming the ILCDirac servers, we applied small test production with updating ILCDirac client version 
 and small issues have observed but quickly fixed.
 Meanwhile, we reproduced some forgotten samples; single particles, LCFI calibration, and higgs samples with semi-digital hadron calorimeter option (ILD\_l5\_o2\_v02).
 After confirming the production workflow with updated ILCDirac, 
 production has been resumed and $2f$ large cross section channel production is launched.
 

 \section{Conclusion}
 We have conducted mass production of new 250 GeV MC samples fully utilizing the ILCDirac system since autumn 2020.
 According to the initial plan, more than 2 Giga events of about 240 processes are processed creating about 130 TB of DST files for high statistics physics studies.
 By the end of 2021 April, the production of all initial processes has been completed except $2f$ hadronic channels. 
It is expected to take a couple of more months to be completed remaining $2f$ channels.
In order to speed up, the ILDProd parameters are being tuned for larger number of concurrent jobs with lower job failure rate.
Additional channels, such as $6f$ and $ee\gamma$, will be produced when generator samples are ready to use.
 
 \section*{Acknowledgments}
 
We would like to thank ILCDirac administrators, especially Andre Sailer, 
 for the support and maintenance of ILCDirac system
 and ILD software group conveners for the continuous support and discussion regarding the production.
 We would also acknowledge all the computing sites who provide the resources and support for ILC Virtual Organization.

  \begin{footnotesize}
  
 \end{footnotesize}


\end{document}